\documentclass[onecolumn,draft, 11pt]{IEEEtran}

\usepackage{color}

\usepackage{graphicx,tabularx,array,amsmath,amsthm,thmtools}

\usepackage{mathtools}

\usepackage{amsfonts}
\usepackage{bm}
\usepackage{bbm}
\usepackage{makecell}
\usepackage{multirow}
 \usepackage{amssymb}
\usepackage{txfonts}
\usepackage[T1]{fontenc}
\usepackage{tikz}
\usepackage[scr=dutchcal]{mathalfa}
\usepackage{ textcomp }
\usepackage{enumitem}

\usepackage{cite}
\newcommand\blfootnote[1]{%
  \begingroup
  \renewcommand\thefootnote{}\footnote{#1}%
  \addtocounter{footnote}{-1}%
  \endgroup
}
\newtheorem{Theorem}{Theorem}
\newtheorem{Proposition}{Proposition}

\newtheorem{Definition}{Definition}

\theoremstyle{definition}

\begin{document}

\title{Tradeoff Between Delay and High SNR
Capacity in Quantized MIMO Systems
}

\author{Abbas Khalili$^1$, Farhad Shirani$^1$, Elza Erkip$^1$, Yonina C. Eldar$^2$\\
$^1$Dept. of Electrical and Computer Engineering,
New York University, NY. \\
$^2$ Dept. of Electrical Engineering, Technion Institute of Technology, Israel\date{} }


\maketitle
\begin{abstract}

Analog-to-digital converters (ADCs) are a major contributor to the power consumption of multiple-input multiple-output (MIMO) communication systems with large number of antennas. Use of low resolution ADCs has been proposed as a means to decrease power consumption in MIMO receivers. However, reducing the ADC resolution leads to performance loss in terms of achievable transmission rates.
In order to mitigate the rate-loss, the receiver can perform analog processing of the received signals before quantization. 
Prior works consider one-shot analog processing where at each channel-use, analog linear combinations of the received signals are fed to a set of one-bit threshold ADCs. 
In this paper, a receiver architecture is proposed which uses a sequence of delay elements to allow for blockwise linear combining of the received analog signals. In the high signal to noise ratio regime, it is shown that the proposed architecture achieves the maximum achievable transmission rate given a fixed number of one-bit ADCs.
Furthermore, a tradeoff between transmission rate and the number of delay elements is identified which quantifies the increase in maximum achievable rate as the number of delay elements is increased.  \blfootnote{This work is supported by NYU WIRELESS Industrial Affiliates and
National Science Foundation grant SpecEES-1824434.}
%It is shown that for a fixed number of delay elements, application of quantizers with non-zero thresholds leads to improved rates compared to zero threshold quantizers. However, these gains vanish as the number of delay elements is increased asymptotically.  
\end{abstract}

\section{Introduction}
One of the most significant challenges in the development of 5G cellular communication technologies is energy consumption. The use of large antenna arrays leads to energy demands which are inconsistent with the limited power budget available in mobile devices and small-cell access points \cite{rangan2014millimeter}.  Analog to digital converters (ADCs) are a major contributor to the power consumption in multiple-input multiple-output (MIMO) receivers.  In conventional MIMO systems with digital beamforming, it is assumed that each receiver antenna is connected to a high resolution ADC \cite{DigBF1}. In standard ADC design, the power consumption is proportional to the number of quantization bins and hence grows exponentially in the number of output bits \cite{walden1999analog,eldar2015sampling}.
%Application of high resolution ADCs is especially costly when using channels with large bandwidths since the power consumption of an ADC grows linearly in bandwidth \cite{BR}. 
One method which has been proposed to address high power consumption in MIMO systems with large number of antennas is to use low resolution ADCs (e.g. one-bit threshold ADCs) at each receiver antenna \cite{mo2014channel,mezghani2012capacity,MIMO1,abbasISIT2018,rini2017generalITW,mo2015capacity,alkhateeb2014mimo,koch2013low}. Reducing the ADC resolution decreases power consumption, however, it also results in lower transmission rates. This suggests a tradeoff between transmission rate and power consumption which is controlled by the  number and resolution of the ADCs at the receiver.

In classical information theory, it is well-known that in order to achieve optimal transmission rates, communication must be performed over asymptotically large blocks of data \cite{csiszar2011information}. More precisely, an optimal decoder performs a possibly non-linear operation on an asymptotically large block of channel outputs. In MIMO systems using high resolution ADCs, the discretization loss is negligible due to the fine quantization grid. Simultaneous blockwise decoding is made possible by storing the digital output and performing the decoding operation over large blocklengths in the digital domain. However, when low resolution ADCs are used, discretizing the individual channel outputs prior to blockwise decoding leads to loss of information  and suboptimal performance \cite{mezghani2012capacity}.  In particular, restricting to one-bit ADCs leads to large quantization noise, and a significant reduction in achievable rates \cite{mo2015capacity,abbasISIT2018}.

\begin{figure}[t]
\centering 
\includegraphics[width=0.7\textwidth,draft=false]{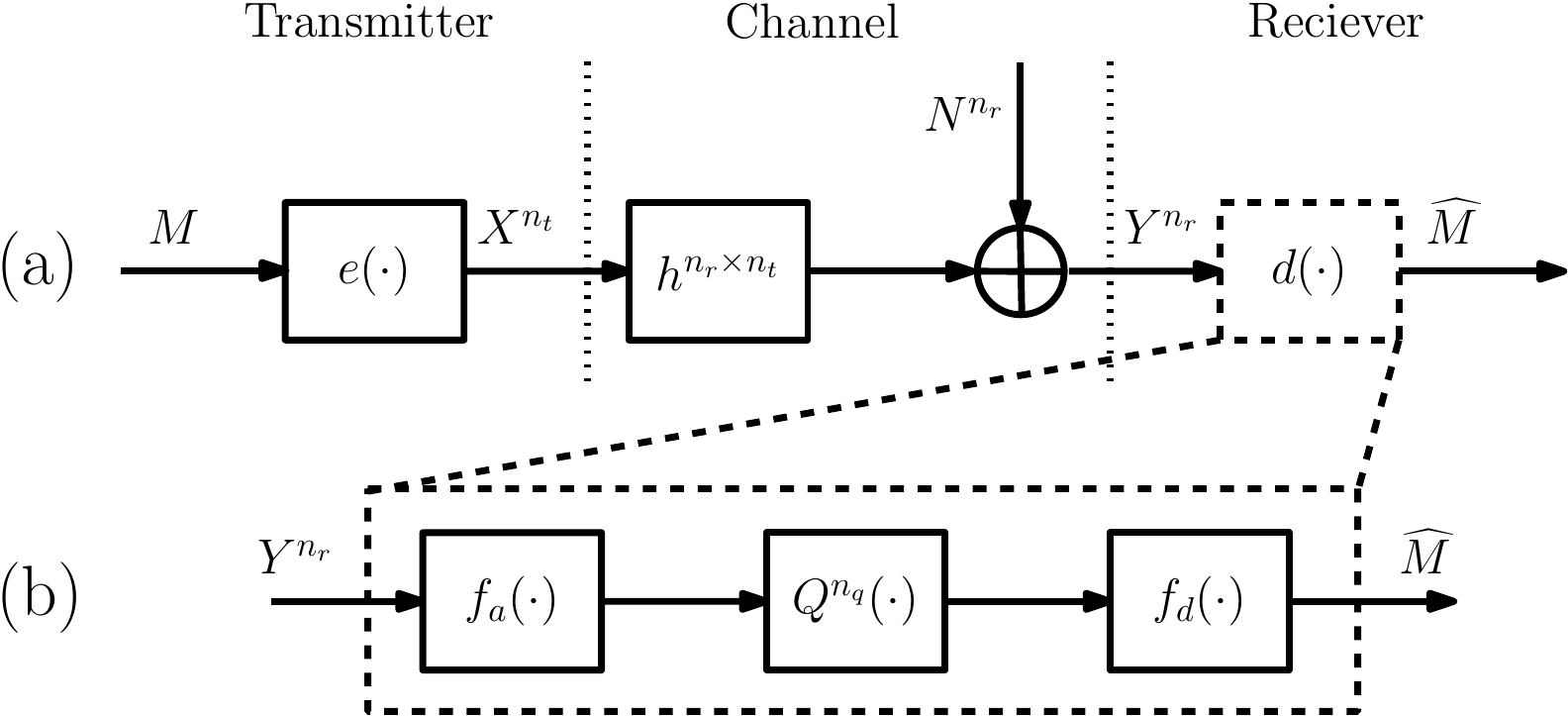}
\caption{The top figure shows a MIMO channel with $n_t$ transmit antennas and $n_r$ receive antennas. The bottom figure is the receiver architecture consisting of analog processing module $f_a(\cdot)$, $n_q$ one-bit ADCs $Q^{n_q}(\cdot)$, and a digital processing module $f_d(\cdot)$. The function $f_a(\cdot)$ may have causal memory. 
}
\label{fig:classic}
\end{figure}

Rate-loss due to low resolution quantization can be {attributed to} 
%broken down into 
two constituents which we call \textit{intrinsic} and \textit{extrinsic} rate-loss. To elaborate, consider the MIMO communication system shown in Fig. \ref{fig:classic}. Assume that the receiver is equipped with $n_q$ one-bit threshold ADCs.
An upper-bound on the channel capacity is given by $\min(n_q,C)$ bits per channel use, where $C$ is the capacity of the MIMO channel when using ADCs with very high resolution. In other words, due to  the restriction on the number of ADCs, the channel capacity is decreased by at least $C-\min(n_q,C)$ bits per channel-use. This \textit{intrinsic} rate-loss cannot be reduced by improving the receiver architecture design without the use of additional one-bit ADCs. 
%The reason is that at each channel-use, the ADCs output at most $n_q$ bits of information. 
Considering the receiver architecture in Fig. \ref{fig:classic}(b), in practice only a limited set of analog operations illustrated as $f_a(\cdot)$ in the figure may be implemented.  Prior works have studied the use of one-shot analog linear combiners and threshold ADCs \cite{mo2015capacity,abbasISIT2018,rini2017generalITW, alkhateeb2014mimo,koch2013low}.
It has been shown that the maximum rate achievable using the architecture in Fig. \ref{fig:classic} is less than $\min(n_q,C)$ due to practical limitations in analog processing \cite{abbasISIT2018}.  More precisely, the communication system suffers an additional \textit{extrinsic} rate-loss of $\min(n_q,C)-R^*$ bits per channel-use, where $R^*$ is the maximum achievable rate when these practical limitations are taken into account. In theory, the extrinsic rate-loss may be reduced by improving the receiver architecture design.

 In this work, we consider communication over MIMO channels where one-bit threshold ADCs are used at the receiver. We propose a blockwise analog processing module in which delay elements are used to reduce the extrinsic rate-loss due to one-bit ADCs.
We show that for a large class of MIMO channels, in the high signal to noise ratio (SNR) regime, the proposed architecture completely  eliminates the extrinsic rate-loss and achieves the maximum transmission rate among all receiver architectures with a fixed number of one-bit ADCs. We show the existence of a fundamental tradeoff between the number of delay elements and the maximum achievable rate for the proposed architecture. In addition, we show that given a fixed number of one-bit ADCs and delay elements, non-zero thresholds are necessary to achieve optimal transmission rates; whereas,  using asymptotically large numbers of delay elements leads to optimal rates without requiring non-zero thresholds.

In the receiver architecture proposed in this paper, the ADC thresholds are chosen according to the channel gain matrix and are assumed to be fixed throughout the transmission block.
In a companion paper \cite{arxiv_Multi_ADC_2018}, we propose a class of adaptive threshold receiver architecture, where the quantization thresholds at each channel-use are dependent on the channel outputs in the previous channel-uses. The fixed threshold architecture in this paper is simpler to implement. However, the adaptive threshold architecture in \cite{arxiv_Multi_ADC_2018} is more amiable to analysis for point-to-point (PtP) communication in the low SNR regime and multiterminal communications.

The rest of the paper is organized as follows.  Section \ref{sec:System Model} describes the system model. 
Section \ref{sec:arch} includes the proposed receiver architecture along with an analysis of the resulting achievable rate region. 
Section \ref{sec:conclusion} concludes the paper.

{\em Notation:} the random variable $\mathbbm{1}_{\mathcal{E}}$ is the indicator function of the event $\mathcal{E}$.
 The set of numbers $\{1,2,\cdots, n\}, n\in \mathbb{N}$ is represented by $[n]$. 
 For a given $n\in \mathbb{N}$, the $n$-length vector $(x_1,x_2,\hdots, x_n)$ is written as $x^n$. The subvector $(x_k,x_{k+1},\cdots,x_n)$ is denoted by $x_k^n$. We write $||x^n||_2$ to denote the $L_2$-norm of $x^n$. An $n\times m$ matrix is written as $h^{n\times m}=[h_{i,j}]_{i,j\in [n]\times [m]}$,
% $h^{n\times m}=[h^{n}(1), h^{n}(2), \cdots, h^{n}(m)]$. 
and $[h^{n\times m}]^{\dagger}$ is the transpose of ${h^{n\times m}}$. Let $h^{n_t}_{(i)}, i\in [m]$ be a sequence of column vectors; the notation $[h^{n}_{(1)}, h^{n}_{(2)}, \cdots, h^{n}_{(m)}]^\dagger$ represents the column vector of length $m n$ consisting of the concatenation of the original vectors. The $n\times n$ identity matrix is shown by $I_n$.  
 We write $a^{n\times m}\otimes b^{n'\times m'}$ to denote the Kronecker product of matrices. 
 % The vector $(x_1,x_2,\cdots, x_n)$ is written as $x^n$. The $m\times t$ matrix $[g_{i,j}]_{i\in [m], j\in [t]}$ is denoted by $g^{m\times t}$.
 %For a random variable $X$, the corresponding probability space is $(\mathcal{X}, \mathbf{F}_{X}, P_X)$, where $\mathbf{F}_X$ is the underlying $\sigma$-field. The set of all subsets of $\mathcal{X}$ is written as $2^{\mathcal{X}}$. For an event $\mathcal{A}\in 2^{\mathcal{X}}$, the random variable $\mathbb{1}_{\mathcal{A}}$ is the indicator function of the event.  
The value of $i$ modulo $k$ is represented by $mod_k(i), i,k\in \mathbb{N}$. 
%A random process (signal) is represented by $\{X(i), i\in \mathbb{N}\}$, where $X(i)$ is the  value of the signal at time $i$. 
The binary entropy function is $h_b(x) = -x\log{x} -(1-x) \log(1-x)$. %{\color{red} Abbas: ask Farhad.}
	%
%High {\rm SNR} results are given in Sec. \ref{sec:High {\rm SNR} Results}. 
	%
%Finally, Sec. \ref{sec:Proposed Research Direction} contains some prospect for the future research.
	%

\section{System Model and Preliminaries}
\label{sec:System Model}

\subsection{System Model}
We consider a PtP communication system  characterized by the triple $(n_t,n_r, h^{n_r\times n_t})$, where $n_t$ is the number of transmitter antennas, $n_r$ is the number of receiver antennas, and $h^{n_r\times n_t}$ is the channel gain matrix. The matrix $h^{n_r\times n_t}$ is assumed to be fixed over the transmission block, and known at the transmitter and receiver.
The channel input and output vector pair $(X^{n_t}, Y^{n_r})$ is related through 
\begin{equation}
Y^{n_r}=h^{n_r\times n_t}X^{n_t}+N^{n_r},    
\label{eq:channel}
\end{equation}
where $N^{n_r}$ is a vector of independent and identically distributed Gaussian variables with unit variance and zero mean, and the channel input has average power constraint $P$. It is assumed that $n_q$ one-bit threshold ADCs are available at the receiver. The receiver uses the architecture shown in Fig. \ref{fig:classic}(b) which consists of an analog signal processing step prior to quantization and a digital signal processing step afterwards. The channel output is processed in the analog domain and the resulting vector is input to the ADCs. The output of the ADCs is processed in the digital domain to reconstruct the message. In its most general form, the analog processor may have causal memory. More precisely, the output of $f_a(\cdot)$ at time $i$, may depend on the matrix of received channel outputs $Y^{i\times n_r}$, where the $j$th row of $Y^{i\times n_r}$ is the channel output at time $j, j\leq i$. Let $n\in \mathbb{N}$ be the length of the transmission block and define
$\mathcal{G}_a=\{f_a: \mathbb{R}^{n\times n_r} \to \mathbb{R}^{nn_q}\}$ as the space of all functions with causal memory. Due to practical considerations, only a subset of the functions in $\mathcal{G}_a$ are implementable. We denote the space of implementable functions by $\mathcal{F}_a$. The set of implementable functions $\mathcal{F}_a$ which are considered in this paper will be discussed in Section \ref{sec:block}.  
The communication problem is formalized below.
\begin{Definition}[\bf{QMIMO}]
 A PtP MIMO system with one-bit ADCs (QMIMO)  is characterized by the tuple $(n_t,n_r, h^{n_r\times n_t},n_q, {\mathcal{F}_a})$, where $n_q$ is the number of one-bit ADCs, and ${\mathcal{F}_a}\subseteq\mathcal{G}_a$.

Let $n,\Theta\in \mathbb{N}$ be a pair of natural numbers, $f_a\in {\mathcal{F}_a}$ an implementable analog function and ${t}^{nn_q}\in \mathbb{R}^{nn_q}$ a vector of quantization thresholds, where $t_{in_q-n_q+1}^{in_q}, i\in [n]$ is the threshold vector in the $i$th channel use.
An $(n,\Theta,f_a,{t}^{nn_q})$-transmission system consists of a pair of encoding and decoding functions $(e,d)$ where $X^{n\times n_t}=e(M)$ is the channel input over $n$ channel uses and $\widehat{M}=d(Y^{n\times n_r})= f_d(Q^{nn_q}(f_a(Y^{n\times n_r})+ {t}^{nn_q}))$ is the message reconstruction, $f_d: \{0,1\}^{nn_q}\to [\Theta]$ is a vector of Boolean functions, $Q^{nn_q}(x^{nn_q})=(\mathbbm{1}_{\{x_1\geq 0\}},\mathbbm{1}_{\{x_2\geq 0\}},\cdots,\mathbbm{1}_{\{x_{nn_q}\geq 0\}}), x^{nn_q}\in \mathbb{R}^{nn_q}$ is a sequence of one-bit ADCs. Achievability is defined in the standard Shannon sense. The capacity maximized over all implementable analog functions is denoted by $C_{Q}(h^{n_r\times n_t},n_q, {\mathcal{F}_a})$.
% Based on the aforementioned practical constraints, 
% \begin{equation}
%     {\mathcal{F}_a}_n= \{v|~v(Y^{n\times n_r})= v^{nn_q\times nn_r}\tilde{Y}^{nn_r}\}.
%     \label{eq:blockdecoding}
% \end{equation}
%{\color{red} Abbas: above we mentioned that based on practical constraints $v$ can only be linear combiner. would it be better to define $v$ this way here instead of a general formulation?}
\end{Definition}

 In \cite{abbasISIT2018}, a receiver architecture is considered where the analog processing module consists of linear combiners along with non-zero threshold ADCs. The architecture is shown in Fig. \ref{fig:lin_comb}.  The linear combiner matrix $v^{n_q\times n_r}$ is applied to the received signals at each channel-use. This receiver architecture does not allow for temporal processing of the received signals in the analog domain. More precisely, the set $\mathcal{F}_a$ considered in \cite{abbasISIT2018} consists of all memoryless and linear analog processors:
\begin{align*}
    {\mathcal{F}_a}= \{f_a|~f_a(Y^{n\times n_r})= (v^{n_q\times n_r} \otimes I_n) \widetilde{Y}^{nn_r}\},
%    \label{eq:exp:sysISITabbas}
\end{align*} 
where $\widetilde{Y}_{kn+1}^{(k+1)n}, k\in \{0,1,\cdots,n_r-1\}$ is equal to the $k$th row of $Y^{n\times n_r}$.  We call this receiver architecture, \textit{one-shot}. The channel capacity  using this one-shot architecture maximized over all input distributions, threshold vectors, and linear combining matrices is denoted by $C_{OS}(h^{n_r\times n_t},n_q)$.

 It is known that one-shot processing of the analog signals leads to a significant extrinsic rate-loss \cite{abbasISIT2018,mo2015capacity}. In fact, the one-shot capacity is shown to grow at most logarithmically in the number of one-bit ADCs. Consequently, in the high SNR regime, the extrinsic rate-loss due to the application of one-shot receiver architectures is at least\footnote{We write $f(x)=O(g(x))$ if $\lim_{x\to \infty} \frac{f(x)}{g(x)}<\infty$.} $n_q- O(\log{n_q})$ and becomes arbitrarily large as the number of ADCs is increased.

In Section \ref{sec:block}, we introduce blockwise analog processing architectures which use delay elements to allow for temporal processing of the analog signals before quantization. We show in Theorem \ref{th:1} that this allows us to completely eliminate the extrinsic rate-loss in a large class of MIMO systems in the high SNR regime including MIMO systems where the number of one-bit ADCs is at least twice the number of transmitter antennas and receiver antennas. To analyze the performance of the proposed architecture, we utilize a geometric interpretation introduced in \cite{abbasISIT2018,mo2015capacity}. The geometric interpretation is especially helpful in analyzing the set of achievable rates in the high SNR regime. 

The geometric interpretation and combinatorial background is briefly described in the next subsection.

% The function $f_a(\cdot)$ is a linear function characterized by a block-diagonal matrix $v^{n_q\times n_r}\otimes I_n$, where the diagonal blocks are identical copies of the linear combiner matrix $v^{n_q\times n_r}$. The set of analog functions is: 
% \begin{align*}
%     {\mathcal{F}_a}= \{f_a|~f_a(Y^{n\times n_r})= (v^{n_q\times n_r}\otimes I_{n})\widetilde{Y}^{nn_r}, n\in \mathbb{N}\},
% %    \label{eq:exp:sysISITabbas}
% \end{align*}
% where $\widetilde{Y}_{kn+1}^{(k+1)n}, k\in \{0,1,\cdots,n_r-1\}$ is equal to the $k$th row of $Y^{n\times n_r}$. This class of receiver architectures are called \textit{one-shot architectures}. 

%{\color{red} Therefore we have: The set of v consists of linear combiners.\\
%To elaborate more on this we consider the following example which we will reference through out the paper.}
\subsection{Combinatorial Background}
Loosely speaking, as the SNR is increased, the effect of noise in Equation \eqref{eq:channel} becomes negligible and the channel output is almost equal to $h^{n_r\times n_t}X^{n_t}$. In fact, in the absence of noise, the channel output space is $Im(h^{n_r\times n_t})$ the image of the channel gain matrix $h^{n_r\times n_t}$. In the following, we describe the relation between partitions of the subspace $Im(h^{n_r\times n_t})$ and the maximum transmission rate when one-shot receiver architectures are used.
%Let $\overline{\mathcal{R}}= \mathbb{R}^{n_t}/ker(h^{n_r\times n_t})$ be the quotient space of the $n_t$-dimensional space by the null space of the channel matrix $h^{n_r\times n_t}$. In other words, 
%\begin{align*}
%    \overline{\mathcal{R}}= \{ker(h^{n_r\times n_t})+x^{n_t}| x^{n_t}\in \mathbb{R}^{n_t}\}.
%\end{align*}
%The quotient space $\overline{\mathcal{R}}$ is naturally isomorphic to $Image(h^{n_r\times n_t})$ \cite{DM}.
%Let $\phi:\overline{\mathcal{R}} \to Image(h^{n_r\times n_t})$ be the natural isomorphism. 
%In this  interpretation, each one-bit ADC may be viewed as a hyperplane partitioning the Euclidean space $\mathbb{R}^{n_t-dim(ker(h^{n_r\times n_t}))}$ through its action on $\overline{\mathcal{R}}$. The quantization thresholds determine the offset value of the hyperplane. The linear combiner rotates the space and consequently determines the normal vector of the hyperplane.

Consider the one-shot architecture described in Fig. \ref{fig:lin_comb}. For a given channel output vector $y^{n_r}$, let $j_i=Q({w}_i), i\in [n_q]$, where ${w}^{n_q}=v^{n_q\times n_r}y^{n_r}+t^{n_q}$ is the input vector to the one-bit ADCs. The binary vector $j^{n_q}$ is the vector of ADC outputs. The set of all channel output vectors $y^{n_r}$ which result in the ADC output vector $j^{n_q}$ is
\begin{align*}
   \mathcal{B}_{j_1,j_2,\cdots,j_{n_q}}
    = \{y^{n_r}\in Im(h^{n_r\times n_t})| Q(w_i)= j_i, i\in [n_q]
     \}.
\end{align*}
For a given pair  $(t^{n_q}, v^{n_q\times n_r})$, the collection of sets  $\mathsf{B}(t^{n_q}, v^{n_q\times n_r})
= \{\mathcal{B}_{j_1,j_2,\cdots,j_{n_q}}| j_i\in \{0,1\}, i\in [n_q]\}$, is a partition of $Im(h^{n_r\times n_t})$. The number of non-empty partition elements corresponds to the number of messages which can be transmitted reliably as the SNR is taken to be asymptotically large.  Note that for some binary vectors $j^{n_q}$, the set $\mathcal{B}_{j_1,j_2,\cdots,j_{n_q}}$ may be empty. For instance, let $n_t=n_r=1$, $n_q=2$, $h^{n_r\times n_t}=1$,  $v^{n_q\times n_t}=
\begin{bmatrix}
1 \\
    1
    \end{bmatrix}
    $, and $t^{n_q}=\begin{bmatrix}
    0 \\
    0
\end{bmatrix}$. Then $\mathcal{B}_{0,1}= \{y| Q(y)=0, Q(y)=1\}=\varnothing$,
similarly, $\mathcal{B}_{1,0}=\varnothing$. 
As a result, the number of partition elements may be less than $2^{n_q}$. In order to increase the transmission rate, it is desirable to choose $(t^{n_q}, v^{n_q\times n_r})$ such that the number of non-empty partition elements is maximized. 

    We use the following proposition throughout the paper. 
%     It is well-known that the number is maximized when $(t^{n_q}, v^{n_q\times n_r})$ satisfies the 
%  \textit{general position} property. 
\begin{Proposition}(\hspace{-0.002in}\cite{winder1966partitions})
\label{Prop:Partition}
The maximum number of non-empty partition elements is given by
\begin{align}
\label{eq:M_th}
    \max_{t^{n_q}, v^{n_q\times n_r}} |\mathsf{B}(t^{n_q}, v^{n_q\times n_r})-\{\varnothing\}|
    = \sum_{i=0}^{rank(h^{n_r\times n_t})} {n_q \choose i}.
\end{align}
Additionally , if the threshold vector is taken to be the all-zero vector, then:
\begin{align}
\label{eq:M_zth}
   \max_{ v^{n_q\times n_r}}|\mathsf{B}(0^{n_q}, v^{n_q\times n_r})-\{\varnothing\}|
    = 2\sum_{i=0}^{rank(h^{n_r\times n_t})-1} {n_q-1 \choose i}.
\end{align}
\end{Proposition}
 The maximum number of non-empty partition regions grows exponentially in $n_q$ since $\log{{n \choose k}}= nh_b(\frac{k}{n})+ O(\log{n})$ as shown in Proposition \ref{Prop:nchoosek}.
 
% \begin{Remark}
% It can be shown that for any given PtP-QMIMO setup, there exists a set of $(t^{n_q}, v^{n_q\times n_r})$ which are in general position, and   $(0^{n_q}, v^{n_q\times n_r})$ in zero-thereshold general position \cite{alexanderson1978simple,winder1966partitions}. Furthermore, $(t^{n_q}, v^{n_q\times n_r})$ pairs which are in general position result in partitions of $Image(h^{n^r\times n^t})$ with the maximum possible number of non-empty partition elements.
% \label{rem:gen_pos}
% \end{Remark}

\section{Blockwise Receiver Architectures}
\label{sec:arch}
\begin{figure}[t]
\centering
\includegraphics[width=0.8\textwidth,draft=false]{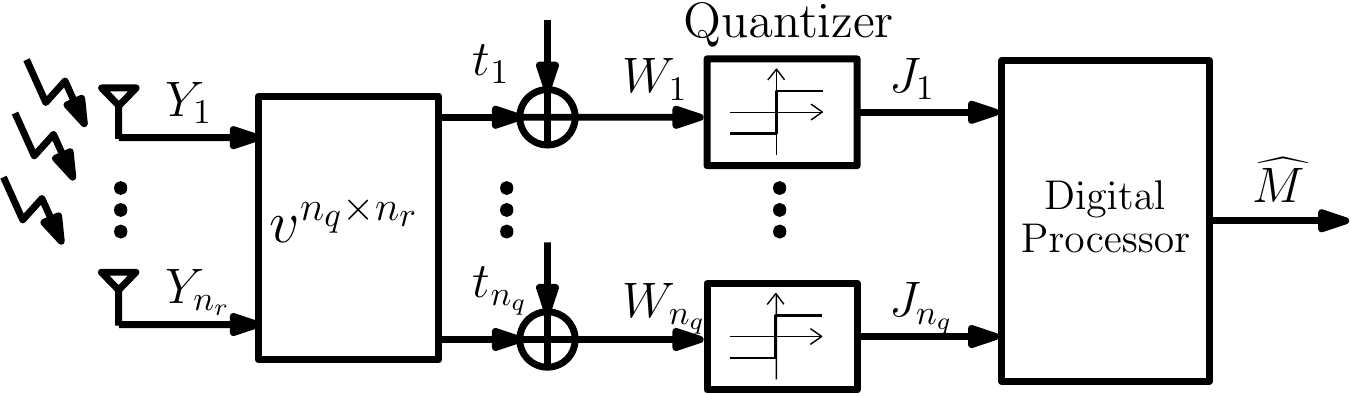}
\caption{A one-shot receiver architecture, where
the linear combiner is characterized by the matrix $v^{n_r\times n_q}$, and the ADC thresholds are $t^{n_q}=(t_1,t_2,\cdots,t_{n_q})$.}
\label{fig:lin_comb}
\end{figure}
\begin{figure}
    \centering
    \includegraphics[width = 4.8in, height= 2.2in ,draft=false]{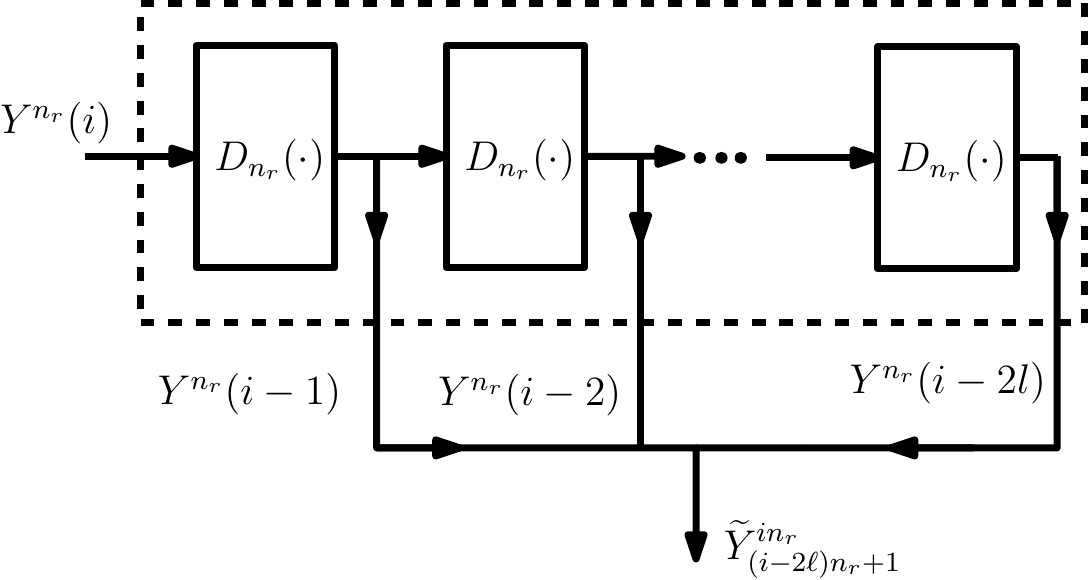}
    \caption{Delay network in the $i$th channel-use where $\widetilde{Y}^{i n_r}_{(i-2\ell)n_r+1} = (Y^{n_r}(i-2\ell+1), Y^{n_r}(i-2\ell+2),\cdots,Y^{n_r}(i-1), Y^{n_r}(i))$ and $i\geq 2\ell$.}
    %which is capable of preserving a block of length $n$ for $n$ time-slots.}
    \label{fig:delay net}
\end{figure}

 We propose \textit{blockwise} receiver architectures in which delay elements are used to perform blockwise temporal processing of the received signals before quantization. Communication is performed in $n=\ell b$ channel-uses, where $n$, $\ell$ and $b$ are called the \textit{blocklength}, \textit{inner blocklength}, and \textit{outer blocklength}, respectively.  The blockwise receiver architecture uses a delay network consisting of $2\ell$ delay elements as shown in Fig. \ref{fig:delay net}, where each delay element $D_{n_r}(\cdot)$ takes the vector of received signals at the $i$th channel-use $Y^{n_r}(i)$ and outputs $Y^{n_r}(i-1)$. In other words, $D_{n_r}(\cdot)$ delays the received analog vector by one channel-use. The stored analog signals are combined using the linear combining matrix $v^{\ell n_q\times \ell n_r}$ over $\ell$ channel-uses.

To clarify the linear combination process, let us describe the first $3\ell$ channel-uses. In the first $\ell$ channel-uses the received signals $Y^{n_r}(i), i\in [\ell]$ are stored in the delay network. In the next $\ell$ channel-uses, the second batch of received signals $Y^{n_r}(i), \ell+1\leq i\leq 2\ell$ are stored in the delay network while the linear combiner 
operates on the previously stored signals $Y^{n_r}(i), i\in [\ell]$. More precisely, for the $i$th channel-use where $\ell+1\leq i\leq 2\ell$, the linear combiner outputs $\overline{Y}_{(i-\ell-1)n_q+1}^{(i-\ell)n_q}$, where $\overline{Y}^{\ell n_q}= v^{\ell n_q\times \ell n_r}\widetilde{Y}^{\ell n_r}$, and 
\[\widetilde{Y}^{\ell n_r} = (Y^{n_r}(1), Y^{n_r}(2),\cdots,Y^{n_r}(\ell-1), Y^{n_r}(\ell)).\]
In the third $\ell$ channel-uses, the third batch of received signals $Y^{n_r}(i), 2\ell+1\leq i\leq 3\ell$ are stored in the delay network while the linear combiner operates on the previously stored signals $Y^{n_r}(i), i\in [\ell]$. This process continues for $b$ blocks of length $\ell$ until the $n$th channel-use, where $n$ is the blocklength. The output of the linear combiner is given to the $n_q$ one-bit threshold ADCs. The threshold vector used in the one-bit ADCs changes periodically with a period of $\ell$ channel-uses. More precisely, let $t^{\ell n_q}\in \mathbb{R}^{\ell n_q}$ and define $\widetilde{t}^{n_q}(k)=t_{kn_q-n_q+1}^{kn_q}, k\in [\ell]$. For the $\ell$ first channel-uses, the threshold vector $\widetilde{t}^{n_q}(i)$ is used in the $i$th channel-use. For the second $\ell$ channel-use, the threshold vector $\widetilde{t}^{n_q}(i-\ell)$ is used in the $i$th channel-use. Generally, for $i\in [n]$, let $k= mod_{\ell}(i)$, the vector $t^{n_q}(k)$ is used as the threshold vector for the one-bit ADCs at the  $i$th channel-use. 

We call the resulting communication system a D-QMIMO system, where $D$ refers to delay. The set of implementable analog functions for this architecture is:
\begin{align*}
    {\mathcal{F}_a}^{\ell}= \{f_a|~f_a(Y^{n\times n_r})= (v^{\ell n_q\times \ell n_r} \otimes I_b) \widetilde{Y}^{nn_r} , b\in \mathbb{N}\},
%    \label{eq:exp:sysISITabbas}
\end{align*}
where \[\widetilde{Y}^{j n_r}_{(j-2\ell)n_r+1} = (Y^{n_r}(j-2\ell+1), Y^{n_r}(j-2\ell+2),\cdots,Y^{n_r}(j-1), Y^{n_r}(j)),\]
where $2\ell\leq j\leq n$. 
The channel capacity optimized over all analog combining matrices, and threshold vectors is denoted by $C_{\ell}(h^{n_r\times n_t},n_q)$ for a given delay $\ell$. 
%{\color{red} Abbas: this is not right? still searching for inner block definition}
% \begin{Definition}[\bf{D-QMIMO}]
% A PtP-QMIMO with delay elements (D-QMIMO) is characterized by $(\ell,n_t,n_r, h^{n_r\times n_t},n_q,{\mathcal{F}_a}^{\ell})$, where: 
% %$n$ number of delay elements, and: 
% \begin{align*}
%     {\mathcal{F}_a}^{\ell}= \{f_a|~f_a(Y^{n\times n_r})= (v^{\ell n_q\times \ell n_r} \otimes I_b) \widetilde{Y}^{nn_r} , b\in \mathbb{N}\},
% %    \label{eq:exp:sysISITabbas}
% \end{align*}
% where $n = \ell b$ is the blocklength. 
% \end{Definition}

%{\color{red} maybe it is better to use the discussed notation b and n one for the number of blocks in the first two definitions and the other for the length of the block in this definition.}

%\begin{Remark}
Note that D-QMIMO systems are a special class of QMIMO systems where the analog processing is restricted to linear operations.
Furthermore, the one-shot setup described in Fig. \ref{fig:lin_comb} is a special case of  D-QMIMO where the length of each inner-block is equal to one (i.e. $\ell=1$). 
%number of delay elements is
As a result, 
$$C_{OS}(h^{n_r\times n_t},n_q)=C_{1}(h^{n_r\times n_t},n_q)\leq C_{\ell}(h^{n_r\times n_t},n_q),$$
where $\ell \geq 2$.
%\end{Remark}

% \begin{Definition}{\bf{Rate Per Quantizer}}
% We define rate per quantization (RPQ) as the ratio of achievable rate over the number of quantizers.
% \end{Definition}

%Define rate per quantization (RPQ) as the value of the rate per quantization
\label{sec:block}
%In our design, the receiver uses delay elements to preserve and decode inner-blocks. More precisely, we store a block of channel outputs using a  number of delay elements which in turn enables joint decoding.
%by utilizing delay elements, we provide a way of storing the channel outputs so that the decoder can process them jointly. 
%The decoder we consider uses delay elements to preserve and decode the transmission inner-blocks. More precisely, the delay elements are used to store the transmitted inner-block until the receiver decodes its message.
%A \textit{delay element} is a circuit in which the output signal vector is delayed for a predetermined duration (here one time-slot) with respect to the input signal vector. A delay element can be built by deploying delay lines \cite{analui2003statistical}. 
%Using the delay elements, we can construct the \textit{delay network} in Figure \ref{fig:delay net}. The considered delay network is composed of $2\ell$ delay elements, each with input and output size of $n_r$. Therefore, this delay network is capable of preserving a complete inner-block for a duration of $\ell$ time-slots. %{\color{red} Abbas: explain the decoder process here.}

% \begin{figure}[t]
%     \centering
%     \includegraphics[scale = 0.47]{figure-decoder-in-kth-time-slot.pdf}
%     \caption{Proposed receiver architecture at the $i$th time-slot where $k =mod_{\ell}(i)$, DN represent delay network, and AC denotes analog combiner.}
%     \label{fig:decoder-kth-time-slot}
% \end{figure}

We derive the bounds provided in Theorem \ref{th:1} below on the performance of the following proposed coding strategy. Consider a D-QMIMO communication system where $(t^{\ell n_q},v^{\ell n_q\times \ell n_r})$ are taken so that the partition $\mathsf{B}(t^{\ell n_q},v^{\ell n_q\times \ell n_r})$ has the maximum number of non-empty elements as described in Proposition \ref{Prop:Partition}. In other words, $(t^{\ell n_q},v^{\ell n_q\times \ell n_r})$ are chosen such that the number of non-empty partitions is equal to $\sum_{i=0}^{\ell rank(h^{n_r\times n_t})} {\ell n_q \choose i}$. We use the fact that $\log{{n \choose k}}= nh_b(\frac{k}{n})+ O(\log{n})$  as shown in Proposition \ref{Prop:nchoosek} and perform a second order analysis of the number of non-empty partition elements as the number of delay elements $\ell$ is increased asymptotically to characterize the set of achievable rates for high SNRs. We make use of the following proposition. 

\begin{Proposition}
\label{Prop:nchoosek}
Let $n\in \mathbb{N}$ and $\lambda\in (0,1)$ such that
\begin{align}
\label{eq:alpha}
    \frac{1}{2}\left(1-\sqrt{1-\frac{1}{3n}} \right)< \lambda
    < \frac{1}{2}\left(1+\sqrt{1-\frac{4}{12n+1}}\right),
\end{align}
Then, the following equality holds:
\begin{align}
\label{Eq:nchoosek}
\log{{n \choose \lambda n}}= n h_b(\lambda)
-\frac{1}{2}\log{n}
-\frac{1}{2}\log{2\pi\lambda(1-\lambda)}
+O\left(\frac{1}{n}\right).
\end{align}
Particularly, for asymptotically large $n$, Equation \eqref{Eq:nchoosek} holds for any fixed $\lambda\in (0,1)$. Furthermore, for $\lambda\in (0,\frac{1}{2})$, we have\footnote{We write $f(x)=o(g(x))$ if $\lim_{x\to \infty}\frac{f(x)}{g(x)}=0$.}
\begin{align}
\label{Eq:nchoosek2}
\log{\sum_{i=0}^{\lambda n}{n \choose i}}= n h_b(\lambda)
-\frac{1}{2}\log{n}
-\frac{1}{2}\log{\frac{2\pi\lambda(1-2\lambda)^2}{1-\lambda}}
+o\left(1\right).
\end{align}

\end{Proposition}
The proof is provided in the Appendix.
\begin{Theorem}
\label{th:1}
For the D-QMIMO communication system with $n_q$ one-bit ADCs, the capacity $C_{\ell}$ satisfies the following
\begin{align*}
n_q h_b(\alpha)-\frac{1}{2\ell}\log{\ell}
-\frac{\mathbbm{1}_{\{\alpha\neq \frac{1}{2}\}}}{2\ell}&\log{\frac{2\pi n_q\alpha(1-2\alpha)^2}{1-\alpha}}
+o\left(\frac{1}{\ell}\right)
\leq C_{\ell}
\\&\leq 
n_q h_b(\beta)-\frac{1}{2\ell}\log{\ell}
-\frac{\mathbbm{1}_{\{\beta\neq \frac{1}{2}\}}}{2\ell}\log{\frac{2\pi n_q\beta(1-2\beta)^2}{1-\beta}}
+o\left(\frac{1}{\ell}\right),
\end{align*}
as {\rm SNR}$\to \infty$, where $\alpha= \min \{\frac{rank(h^{n_r\times n_t})}{n_q},\frac{1}{2}\}$, $\beta=\min\{\frac{n_r}{n_q},\frac{1}{2}\}$.
 Particularly, if $rank(h^{n_r\times n_t})=n_r$, then 
 \[C_{\ell} \to n_q h_b(\beta)\text{ as }\ell\to \infty.\]  
\end{Theorem}
The proof is provided in the Appendix, where a general coding strategy for arbitrary SNRs is presented. The resulting rate is analyzed in the high SNR regime. 

\begin{figure}
    \centering
    \includegraphics[scale = 0.4,draft=false]{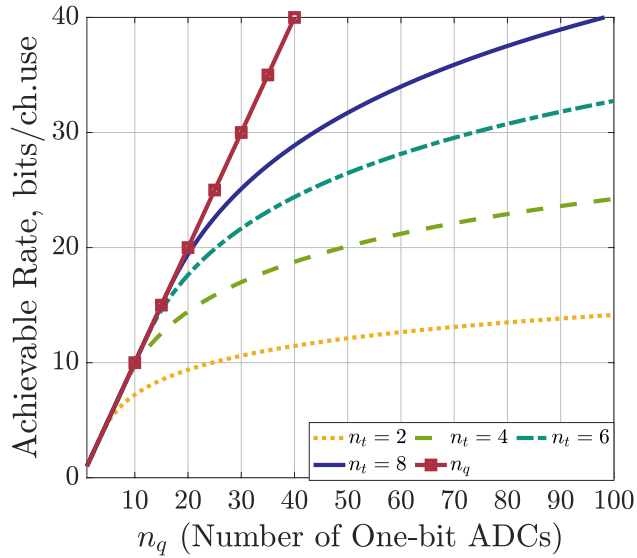}
    \caption{The figure shows the maximum achievable high SNR rate when the number of delay elements $\ell$ is taken to be asymptotically large for the MIMO system with $n_r=10$ and $n_t\in \{2,4,6,8\}$. The red full line is the $R=n_q$ line which is achievable if $n_t, n_r$ are asymptotically large. The channel is assumed to be full-rank. }
    \label{fig:infinite Delay}
\end{figure}

The following observations follow from Theorem \ref{th:1}:
\\\noindent {\bf I)} The capacity approaches $n_q$ as $\ell \to \infty$ if $n_q \leq 2rank(h^{n_r\times n_t})$. Consequently, the extrinsic rate-loss is completely eliminated. This is in contrast with prior works (e.g. \cite{abbasISIT2018}, \cite{mo2015capacity}), where the high SNR capacity grows logarithmically in $n_q$.
\\\noindent {\bf II)} The maximum achievable rate due to using non-zero threshold ADCs is $\frac{1}{\ell} \log{\sum_{i=0}^{\ell rank(h^{n_r\times n_t})} {\ell n_q \choose i}}$, whereas when zero threshold ADCs are used, the maximum rate is $\frac{1}{\ell} \log{\sum_{i=0}^{\ell rank(h^{n_r\times n_t})-1} 2{\ell n_q-1 \choose i}}$. The two values converge to each other as $\ell\to \infty$. This shows that when long delays $\ell$ can be tolerated,  zero threshold ADCs can be used in scenarios where non-zero thresholds are costly to implement without any loss in transmission rate.
\\\noindent {\bf III) }For a fixed number of transmitters $n_t$ and receivers $n_r$, as the number of one-bit ADCs $n_q$ is increased, the maximum achievable rate increases linearly when $n_q\leq 2rank(h^{n_r\times n_t})$ since $h_b(\alpha)=h_b(\frac{1}{2})=1$. The maximum achievable rate increases logarithmically when $n_q\gg 2rank(h^{n_r\times n_t})$ since
\begin{align*}
    n_qh_b\left(\frac{rank(h^{n_r\times n_t})}{n_q}\right)=
    rank(h^{n_r\times n_t})(\log{n_q}- O(\log{n_q})).
\end{align*}
This is shown in Fig. \ref{fig:infinite Delay}, where for a MIMO system with $n_r=10$, the achievable rate in Theorem \ref{th:1} is plotted as a function of $n_q$ for $n_t\in \{2,4,6,8\}$ as the number of delay elements is taken to be asymptotically large.

% \begin{Remark}
% Note that in the coding strategy described in the Appendix, the decoder decodes the message after the $b+1$ blocks of length $\ell$, whereas the blocklength is $n=b\ell$. Let $R$ be the desired rate of transmission. The resulting transmission rate $\frac{b+1}{b}R$ converges to $R$ as $b\to \infty$. The decoding process imposes a constant delay of $\ell$ channel-uses.  
 %\end{Remark}
%\begin{Remark}

%\end{Remark}
 %\begin{Remark}

 %\end{Remark}

%  \begin{figure}[t]
% \centering
% \begin{tikzpicture}[scale= 0.43]

% \draw [help lines] (0,0) grid [step=1] (10,10);
% \draw (0,0) [->,line width=2 pt]-> (10,0);
% \draw (0,0) [->,line width=2 pt]-> (0,10);

% \node at (-0.6,5) {$R$};
% \node at (10,-0.6) {$n_q$};

% \draw [red, ultra thick] plot [domain=0.1:6,samples=100] (\x,\x);
% \draw [dashed] plot [dotted, domain=0.1:10,samples=10] (\x,\x);
% \draw [red, ultra thick] plot [domain=6:10,samples=100 ] (\x,{ -3*log2(3/\x)- (\x-3)*log2(1-3/\x) });
% \end{tikzpicture}
% \caption{The red curve plots the channel capacity when $rank(h^{n_r\times n_t})=3$ as the number of one-bit ADCs is increased for asymptotically large delay.}
% \label{fig:infinite Delay}
% \vspace{-0.2in}
% \end{figure}

\section{Conclusion}
\label{sec:conclusion}
We have considered point-to-point communication over MIMO systems when a limited number of one-bit ADCs are available at the receiver. We have proposed a receiver architecture which uses a sequence of delay elements to allow for blockwise linear combining of the received analog signals. In the high SNR regime, given a fixed number of one-bit ADCs, we have shown that the proposed architecture achieves the maximum transmission rate among all receiver architectures. Furthermore, we have characterized a tradeoff between transmission rate and the number of delay elements which quantifies the increase in maximum achievable rate as the number of delay elements is increased. In a companion paper \cite{arxiv_Multi_ADC_2018} we propose a class of adaptive threshold architectures analyze their performance in PtP communications in the low SNR regime and broadcast channel communications.

\bibliographystyle{IEEEtran}
\bibliography{ref}
\appendix
\section*{Proof of Proposition \ref{Prop:nchoosek} }
From Stirling's approximation, we have: 
\begin{align*} 
n! = {\sqrt {2\pi n}}\left({\frac {n}{e}}\right)^{n}\left(1+{\frac {1}{12n}}+O\left(\frac{1}{n^2}\right) \right).
\end{align*}
Consequently, we have:
\begin{align*}
    &{n \choose n\lambda} = \frac{n!}{(\lambda n)!((1-\lambda)n)!}
    \\&=\frac{ {\sqrt {2\pi n}}\left({\frac {n}{e}}\right)^{n}\left(1+{\frac {1}{12n}}+O\left(\frac{1}{n^2}\right) \right)}
    { {\sqrt {2\pi \lambda n}}\left({\frac {\lambda n}{e}}\right)^{\lambda n}\left(1+{\frac {1}{12\lambda n}}+O\left(\frac{1}{(\lambda n)^2}\right) \right)
     {\sqrt {2\pi (1-\lambda) n}}\left({\frac {(1-\lambda) n}{e}}\right)^{(1-\lambda) n}\left(1+{\frac {1}{12(1-\lambda) n}}+O\left(\frac{1}{((1-\lambda) n)^2}\right) \right)}
    \\&=
    \frac{1}{\sqrt{2\pi n\lambda(1-\lambda)}}
    \times \frac{1}{\lambda^{n\lambda}(1-\lambda)^{n(1-\lambda)}} 
    \times \frac{1+ \frac{1}{12n} + O\left(\frac{1}{n^2}\right)}{\left(1+ \frac{1}{12n\lambda} + O\left(\frac{1}{n^2}\right)\right)
    \left( 1+ \frac{1}{12n(1-\lambda)} + O\left(\frac{1}{n^2}\right) \right)}
    \\&= \frac{1}{\sqrt{2\pi n\lambda(1-\lambda)}}
    \times \frac{1}{\lambda^{n\lambda}(1-\lambda)^{n(1-\lambda)}} \times \frac{1+ \frac{1}{12n} +O\left(\frac{1}{n^2}\right)}{\left(1+ \frac{1}{12n}\left( \frac{1}{1-\lambda} + \frac{1}{\lambda} \right) + O\left(\frac{1}{n^2}\right) \right)}\\
    & \stackrel{\text{(a)}}{=} 
    \frac{1}{\sqrt{2\pi n\lambda(1-\lambda)}} \times \frac{1}{\lambda^{n\lambda}(1-\lambda)^{n(1-\lambda)}}
    \times \left(1+ \frac{1}{12n} + O\left(\frac{1}{n^2}\right)\right) \left(1- \frac{1}{12n\lambda (1-\lambda)} + O\left(\frac{1}{n^2}\right)\right)
    \\& = \frac{1}{\sqrt{2\pi n\lambda(1-\lambda)}}
    \times \frac{1}{\lambda^{n\lambda}(1-\lambda)^{n(1-\lambda)}} \times \left(1+ \frac{1}{12n} \frac{\lambda(1-\lambda) - 1 }{\lambda(1-\lambda)} + O\left(\frac{1}{n^2}\right)\right),
\end{align*}
where in (a) we have used the Taylor series expansion $\frac{1}{1+x} = 1 - x + O(x^2), x\in (-1,1)$. Note that Equation \eqref{eq:alpha} ensures that $\frac{1}{12n}\left( \frac{1}{1-\lambda}+\frac{1}{\lambda}\right) \in (-1,1)$ holds. Consequently, we have:
\begin{align*}
    \log{{n \choose n\lambda}}&= 
    \frac{-1}{2}\log{{{2\pi n\lambda(1-\lambda)}}}
    - {n\lambda}\log{\lambda}-{n(1-\lambda)}\log{(1-\lambda)} 
    +\log{
    \left(1+ \frac{1}{12n} \frac{\lambda(1-\lambda) - 1 }{\lambda(1-\lambda)} + O\left(\frac{1}{n^2}\right)\right)}
    \\&= nh_b(\lambda)-\frac{1}{2}\log{n}-\frac{1}{2}\log{2\pi\lambda(1-\lambda)}+O(1),
\end{align*}
where in the last equality we have used the Taylor expansion $\ln{(1+x)}=1-x+O(x^2), x\in (-1,1]$ to conclude that $\left(1+ \frac{1}{12n} \frac{\lambda(1-\lambda) - 1 }{\lambda(1-\lambda)} + O\left(\frac{1}{n^2}\right)\right)= -\frac{1}{2}\log{2\pi\lambda(1-\lambda)}+O(\frac{1}{n})$. Note that 
Equation \eqref{eq:alpha} ensures that $\frac{1}{12n} \frac{\lambda(1-\lambda) - 1 }{\lambda(1-\lambda)} \in (-1,1]$ holds. This completes the proof of Equation \eqref{Eq:nchoosek}. 

Next, we prove Equation \eqref{Eq:nchoosek2}. Note that 
\begin{align}
\label{Eq:01}
    {n \choose \lambda n-j}= {n \choose \lambda n}\times
    \frac{\lambda n}{n-\lambda n +1}
    \times
    \frac{\lambda n-1}{n-\lambda n +2}\times 
    \cdots 
    \times
    \frac{\lambda n-j+1}{n-\lambda n +j}, 
    \lambda<\frac{1}{2}, j\in [\lambda n].
\end{align}
Let $\beta_n= n^{\epsilon}$, $\epsilon \ll 1$. We have:
\begin{align}
    \label{Eq:11}
    &\frac{\lambda n-i+1 }{n-\lambda n +i}\leq \frac{\lambda n}{n-\lambda n+i}<\frac{\lambda}{1-\lambda}, i\in [\lambda n],
    \\& \label{Eq:12}
    \frac{\lambda n -i+1}{n-\lambda n +i}\geq 
    \frac{\lambda n-\beta_n+1}{n-\lambda n+\beta_n}, \lambda n -\beta_n \leq i \leq \lambda n.
\end{align}
From Equations \eqref{Eq:01} and \eqref{Eq:11}, we have:
\begin{align}
\label{Eq:31}
    \sum_{i=0}^{\lambda n}{n \choose \lambda n-i}\leq {n \choose \lambda n}\left(1+ \frac{\lambda}{1-\lambda}+\left(\frac{\lambda}{1-\lambda}\right)^2+\cdots \right)= {n \choose \lambda n} \frac{1-\lambda}{1-2\lambda}.
\end{align}
From equations \eqref{Eq:01} and \eqref{Eq:12}, we have:
\begin{align*}
    &\sum_{i=0}^{\lambda n}{n \choose \lambda n-i}\geq
    \sum_{i=0}^{\lambda n- \beta_n}
    {n \choose \lambda n}\left(1+ \frac{\lambda n-\beta_n+1}{n-\lambda n+\beta_n}+\left(\frac{\lambda n-\beta_n+1}{n-\lambda n+\beta_n}\right)^2+\cdots +
    \left(\frac{\lambda n-\beta_n+1}{n-\lambda n+\beta_n}\right)^{\beta_n}
    \right)\\
    &= {n \choose \lambda n} \left(\frac{1-\lambda +\frac{\beta_n}{n}-(1-\lambda+\frac{\beta_n}{n})(\frac{\lambda n-\beta_n+1}{n-\lambda n+\beta_n})^{\beta_n+1}}{1-2\lambda+\frac{1}{n}}\right).
\end{align*}
Note that $(1-\lambda+\frac{\beta_n}{n})(\frac{\lambda n-\beta_n+1}{n-\lambda n+\beta_n})^{\beta_n+1}= o(1)$. Consequently, from using the Taylor series expansion of $\frac{1}{1+x}, x\in (-1,1)$, we have:
\begin{align}
\label{Eq:32}
    &\sum_{i=0}^{\lambda n}{n \choose \lambda n-i}= {n \choose \lambda n}
    \left(\frac{1-\lambda }{1-2\lambda}+o(1)\right).
\end{align}
From Equations \eqref{Eq:31} and \eqref{Eq:32} we get 
\begin{align*}
    \sum_{i=0}^{\lambda n}{n \choose \lambda n-i}= {n \choose \lambda n}
    \left(\frac{1-\lambda }{1-2\lambda}+o(1)\right).
\end{align*}
This along with Equation \eqref{Eq:nchoosek} proves Equation \eqref{Eq:nchoosek2}.

\section*{ Proof of Theorem \ref{th:1}}
To prove the achievability (lower bound on $\mathcal{C}_{\ell}$), we describe an outline of the coding strategy for arbitrary SNRs, where the average transmission power constraint is $\mathbb{E}(||X^{n_t}||_2)\leq P$. The resulting communication rate is analyzed as SNR$\to \infty$. Fix $ \ell$ and $b$, where $b$ is the outer code blocklength.
 
 Consider the pair $(t^{\ell n_q},v^{\ell n_q \times \ell n_r})$ which achieve the maximum number of non-empty sets in Proposition \ref{Prop:Partition}. For a given rate $R$, define $\Theta_n = 2^{n R}$, where $n = b \ell$. The message $M \in [\Theta_n]$ is transmitted over $b+1$ transmission blocks each of $\ell$ symbols, where each symbol is in $\mathbb{R}^{n_t}$. Let 
$\mathsf{B}(t^{\ell n_q}, v^{\ell n_q\times \ell n_r})
    = \{\mathcal{B}_{j^{\ell n_q}}| j_i\in \{0,1\}, i\in [\ell n_q]\}$ be the partition corresponding to the pair $(v^{\ell n_q \times \ell n_r},t^{\ell n_q})$ as defined in Section \ref{sec:System Model}. 
    Define $
    \mathcal{J}=  \{j^{\ell n_q}\in \{0,1\}^{\ell n_q}| \mathcal{B}_{j^{n_q}}\neq \varnothing \}$, 
    and let $\hat{y}^{\ell n_r}_{j^{\ell n_q}}\in \mathcal{B}_{j^{\ell n_q}}, j^{\ell n_q}\in \mathcal{J}$ be a set of representatives for the partition elements.
    We define the input vector corresponding to $\hat{y}^{\ell n_r}_{j^{\ell n_q}}$ as
    \begin{align}
    \label{eq:rep}
       \hat{x}^{\ell n_t }_{j^{\ell n_q}}= argmin_{\hat{y}^{\ell n_r}_{j^{\ell n_q}} = (h^{n_r\times n_t} \otimes I_{\ell})x^{\ell n_t}}
       \|x^{\ell n_t}\|_2,
    \end{align}
    and the cost associated with $\hat{y}^{\ell n_r}_{j^{\ell n_q}}$ as:
    \begin{align*}
        c(\hat{y}^{\ell n_r}_{j^{\ell n_q}})
        = min_{\hat{y}^{\ell n_r}_{j^{\ell n_q}}=(h^{n_r\times n_t} \otimes I_{\ell})x^{\ell n_t}}
       \|x^{\ell n_t}\|_2.
    \end{align*}
    We define the transition probability $P_{{Z}^{\ell n_r}|\widehat{Y}^{\ell n_t}}$ as:
    \begin{align*}
        P_{{Z}^{\ell n_r}|\widehat{Y}^{\ell n_t}} (
        \hat{y}^{\ell n_r}_{{k}^{\ell n_q}}|
        \hat{y}^{\ell n_r}_{j^{\ell n_q}})
        = P\left (v^{\ell n_q\times \ell n_r} (\hat{y}^{\ell n_r}_{j^{\ell n_q}}+N^{\ell n_r})+ t^{\ell n_q}\in \mathcal{B}_{{k}^{\ell n_q}}\right),
    \end{align*}
\noindent where $j^{\ell n_q},~k^{\ell n_q} \in \mathcal{J}$. Let ${C}_{outer}$ be the capacity of the PtP channel with the transition probability $P_{{Z}^{\ell n_r}|\widehat{Y}^{\ell n_r}}$ subject to the average power constraint $\mathbb{E}(c(\widehat{Y}^{\ell n_r}))\leq \ell P$. We first construct a family of capacity achieving codes for this channel using standard random coding methods. Each symbol $\widehat{Y}_{I^{\ell n_q}}^{\ell n_r}$ in the randomly generated codewords has alphabet $\mathbb{R}^{\ell n_r}$. In order to transmit the symbol $\widehat{Y}_{I^{\ell n_q}}^{\ell n_r}$, the transmitter finds the corresponding input $X^{\ell n_t}$ (Equation \eqref{eq:rep}) and transmits the vector over $\ell$ channel uses. As the SNR goes to infinity, the channel $ P_{{Z}^{\ell n_r}|\widehat{Y}^{\ell n_t}}$ becomes noiseless. The resulting rate converges to
\begin{align*}
  \frac{1}{\ell}I(\widehat{Y}^{\ell n_r};Z^{\ell n_r})  \to \frac{1}{\ell} H(\widehat{Y}^{\ell n_r})
&=\frac{1}{\ell} \log{\sum_{i=0}^{\ell rank(h^{n_r\times n_t})} {\ell n_q \choose i}}
\\&=
n_q h_b(\alpha)
+\frac{1}{\ell}\left(-\frac{1}{2}\log{\ell n_q}
-\frac{1}{2}\log{\frac{2\pi\alpha(1-2\alpha)^2}{1-\alpha}}
+o\left(1\right)\right).
\end{align*} 
This completes the proof of achievability.  Next we prove the converse result. Using Fano's inequality, it is straightforward show that $C_{\ell}\leq \frac{1}{\ell} \log{|\widehat{\mathcal{Y}}^{\ell n_r}|}$. Note that $|\widehat{\mathcal{Y}}^{\ell n_r}|$ is equal to the number of partition regions of the $\ell n_r$ dimensional Euclidean space. From Proposition \ref{Prop:Partition}, the maximum number of partitions is equal to ${\sum_{i=0}^{\ell n_r} {\ell n_q \choose i}}$. By the same arguments as in the achivability proof, it can be shown that:
\begin{align*}
   \frac{1}{\ell}\log{\sum_{i=0}^{\ell n_r} {\ell n_q \choose i}}
  =
  n_q h_b(\beta)+
\frac{1}{\ell}\left(-\frac{1}{2}\log{\ell n_q}
-\frac{1}{2}\log{\frac{2\pi\beta(1-2\beta)^2}{1-\beta}}
+o\left(1\right)\right).
\end{align*}

%More precisely, let $(e_{b_i},f_{b_i}), i\in \mathbb{N}$ be a family of encoding and decoding functions with blocklength $b_i$ and rates approaching ${C}_{outer}$ constructed using standard information theoretic methods. 

% Define  $\widehat{Y}^{b \ell n_r}= e_{b_i}(M)$ , and let $X^{\ell n_t}= [X^{n_t}{(1)} X^{n_t}{(2)} \cdots X^{n_t}{(\ell)}]^{\dagger}$ be the input vector corresponding to $\widehat{Y}_{(i-1) \ell n_r+1}^{i\ell n_r}$. At the $k$th channel-use, the transmitter sends $X^{n_t}{(k)}$. The receiver obtains $Y^{n_r}(k)$. The initial $\ell$ transmissions are stored in the delay network, and the receiver feeds $v^{\ell n_q \times \ell n_r}(h^{n_r\times n_t}\otimes I_{\ell})[X^{n_t}{(1)} X^{n_t}{(2)} \cdots X^{n_t}{(\ell)}]^{\dagger}$ to the ADCs
% in the next $\ell$ channel-uses while storing the ensuing $\ell$ received transmissions in the delay network. The receiver proceeds by processing the second $\ell$ received signal vectors in a similar fashion and storing the third $\ell$ received signal vectors in the delay network. The process continues for $b+1$ transmission blocks. The decoding function $d_b(\cdot)$ is used to reconstruct $M$ from the quantized outputs. 
 
\end{document}